\documentclass[aps,nofootinbib, prd,twocolumn,superscriptaddress]{revtex4-2}
\usepackage[utf8]{inputenc}
\usepackage[normalem]{ulem} %I have activated it to be able to cross text, I will comment it later!
\usepackage{amsmath}
\usepackage{amssymb}
\usepackage{bbm}
\usepackage{bm}
\usepackage{color}
\usepackage{physics}
\usepackage{graphicx}
\usepackage{epstopdf}
\usepackage{aligned-overset}
\usepackage{hyperref}
\hypersetup{colorlinks=true,linktoc=page,linkcolor=blue,citecolor=red,urlcolor=cyan}

\newcommand{\colm}[1]{#1}
\usepackage[english]{babel}

\renewcommand{\vec}[1]{\mathbf{#1}}

\newcommand{\mathi}{{\text{i}}}

\newcommand{\dime}{d}

\usepackage{xparse}

\NewDocumentCommand{\dkpara}{O{k} O{\dime}}{\frac{{\rm d} {#1}^{\parallel}}{(2\pi)^{#2}}}
\NewDocumentCommand{\dkd}{O{k} O{\dime}}{ \frac{{\rm d}^{#2} {#1}}{(2\pi)^{#2}}}
\NewDocumentCommand{\dxd}{O{x} O{\dime}}{ {\rm d}^{#2} {#1} }
\NewDocumentCommand{\ad}{O{} O{}}{ \bigl \langle {#1}\bigr\rangle_{{\rm ad}{#2}} }

\makeatletter
\def\@tocline#1#2#3#4#5#6#7{\relax
  \ifnum #1>\c@tocdepth % then omit
  \else
    \par \addpenalty\@secpenalty\addvspace{#2}%
    \begingroup \hyphenpenalty\@M
    \@ifempty{#4}{%
      \@tempdima\csname r@tocindent\number#1\endcsname\relax
    }{%
      \@tempdima#4\relax
    }%
    \parindent\z@ \leftskip#3\relax \advance\leftskip\@tempdima\relax
    \rightskip\@pnumwidth plus4em \parfillskip-\@pnumwidth
    #5\leavevmode\hskip-\@tempdima
      \ifcase #1
       \or\or \hskip 2em \or \hskip 4em \else \hskip 6em \fi%
      #6\nobreak\relax
    \dotfill\hbox to\@pnumwidth{\@tocpagenum{#7}}\par
    \nobreak
    \endgroup
  \fi}
\makeatother
\usepackage{graphicx}
\usepackage{dcolumn}
\usepackage{bm}
\usepackage{xcolor}
\usepackage{amsmath,amssymb,amsfonts,latexsym,cancel}
\usepackage{color}
\usepackage{soul}
\usepackage{hyperref}
\usepackage{slashed}
\usepackage{float}
\usepackage{amsmath}
\usepackage{amsfonts}
\usepackage{amssymb}
\usepackage{mathrsfs}
\usepackage{braket}
\usepackage{physics}

\allowdisplaybreaks

\newcommand{\be}{\begin{equation}}
\newcommand{\ee}{\end{equation}}
\newcommand{\bea}{\begin{eqnarray}}
\newcommand{\eea}{\end{eqnarray}}

\newcommand{\mom}{{\bf k}}
\newcommand{\ope}{\mathcal{Q}}

\begin{document}

\title{Conformally flat gravitational analogues to the Schwinger effect}

\author{S.~A.~Franchino-Viñas}

\affiliation{Departamento de Física, Facultad de Ciencias Exactas, Universidad Nacional de La
Plata, C.C. 67 (1900), La Plata, Argentina}
\affiliation{CONICET, Godoy Cruz 2290, 1425 Buenos Aires, Argentina}
\affiliation{Universit\'e de Tours, Universit\'e d'Orl\'eans, CNRS, Institut Denis Poisson, UMR 7013, Tours, 37200, France
}

\author{F.~D.~Mazzitelli}
\affiliation{Centro At\'omico Bariloche, 
Comisi\'on Nacional de Energ\'\i a At\'omica, R8402AGP Bariloche, Argentina}

\affiliation{
Instituto Balseiro, Universidad Nacional de Cuyo, R8402AGP Bariloche, Argentina. }

\author{S.~Pla}
\affiliation{Physik-Department, Technische Universit\"at M\"unchen, James-Franck-Str., 85748 Garching, Germany}

\begin{abstract}
 We study particle creation for scalar fields in conformally flat spacetimes using resummed heat-kernel techniques. We make use of an analogy between quantum scalar fields in conformally flat spacetimes and scalar field theories with a Yukawa coupling in Minkowski space. The correspondence holds exactly at the level of the effective action and includes nonconformal curvature couplings. This framework provides access to particle creation at strong curvature. 
In a radiation-dominated universe, the particle production rates in arbitrary dimensions are independently confirmed through explicit calculations of the Bogoliubov coefficients. We also find new exact gravitational analogues of the Schwinger effect in quantum field theory in curved spacetime.
\end{abstract}

\maketitle

%%%%%%%%%%%%%%%%%%%%%%%%%%%%%%%%%%%%%%
%%%%%%%%%%%%%%%%%%%%%%%%%%%%%%%%%%%%%%
%%%%%%%%%%%%%%%%%%%%%%%%%%%%%%%%%%%%%%
%%%%%%%%%%%%%%%%%%%%%%%%%%%%%%%%%%%%%%

\section{Introduction}\label{sec:intro}

Intense fields are believed to be ubiquitous in our universe, arising for example  in the early universe and in the vicinity of singularities. In such scenarios, the description of physical phenomena frequently necessitates nonperturbative techniques capable of encompassing the strong-field character of the backgrounds. The archetype of such effects is pair creation, which, to cite a few examples, has  clear  implications for cosmology~\cite{Kolb:2023ydq} and for the evaporation of black holes~\cite{Hawking:1974rv}.

Recently, in Refs.~\cite{Franchino-Vinas:2025ejo, Franchino-Vinas:2023wea} the authors of the present letter have developed a heat-kernel approach to obtain (partially) resummed effective actions, by appropriately integrating out scalar and spinor fields in the presence of electromagnetic and Yukawa backgrounds.
Our analytical approach is valid for an arbitrary number of spacetime dimensions and general backgrounds; moreover, as long as a ``high-intensity'' feature is present in the latter, the results condense into compact formulae. This is in contrast with other available analytic nonperturbative approaches, for which only certain solvable cases are tractable, e.g. when using the Bogoliubov technique~\cite{Parker:2009}  and the analytic worldline instantons~\cite{Dunne:2006st, Torgrimsson:2017pzs}. In general cases, it seems unavoidable to try to resort to numerics, such as the numeric worldline instanton method~\cite{Semren:2025dix}, which is nevertheless so far restricted to simple cases.

Our heat-kernel results can be readily used to compute, in the presence of background fields, the transition probability relevant for the pair creation process, i.e. the vacuum persistence probability.
Indeed, in the in-out formalism, the vacuum is respectively described by the $\vert \text{in}\rangle$ and $\vert \text{out} \rangle$ states at early and late times, so that such a probability is tightly linked to the effective action $\Gamma$ through the formula
\be\label{eq:inout}
\vert\langle \text {out}|\text{in}\rangle \vert^ 2= \vert e^{\mathi \Gamma}\vert^2=:e^{-P}\,,
\ee
where $P$ signals the instability of the vacuum and can be interpreted as  the (total, integrated) probability of pair creation.

In this letter, our goal is to show that the heat-kernel outcomes of Refs.~\cite{Franchino-Vinas:2025ejo, Franchino-Vinas:2023wea} have a wider range of applicability than primarily thought, including even gravitational scenarios. In order to do so,
we are going to appeal to analogue systems.  In brief, we will employ the fact that for a scalar field in a conformally flat geometry, the Weyl-rescaled action in conformal time coincides with the action of a scalar field in Minkowski spacetime with a position-dependent potential (or effective mass), so that our resummed heat-kernel results for Yukawa backgrounds apply. Furthermore, we identify a class of FLRW metrics for which the vacuum persistence probability takes a Schwinger-like form, closely paralleling Scalar Quantum Electrodynamics (SQED) in a constant electric field.

 Analogies between quantum fields in curved spacetime and gauge or scalar systems have of course been developed in the past by means of other techniques.
For instance, in  Refs.~\cite{Haouat:2011uy, Rajeev:2019okd}, a massive, conformally coupled scalar field was considered in a variety of FLRW spacetimes. The corresponding modes have a time-dependent frequency
which was  interpreted in terms of a (to be determined) background electromagnetic field and the pair creation probability was computed in conformal time thanks to the Bogoliubov method.  Instead, the approach of Refs.~\cite{vanSuijlekom:2025tjt} is based on the use of a different time variable, which is neither the cosmological nor the conformal time~\cite{Parker:1976vvi}. In a given time-dependent metric, they employ the Bogoliubov method to compute the pair production probability of  a massless and minimally coupled scalar field.

 Contrary to those cases, our analogies will in principle be applicable beyond FLRW universes and are going to be at the level of the effective action, i.e. not just at the level of the single modes. A special emphasis will be given to radiation-dominated universes, for which we will in parallel derive the  results in the Bogoliubov formalism and for arbitrary dimensions. Moreover, we will highlight the existence of new analogue cosmological evolutions for cases in which pair creation is induced by a  nonconformal coupling to the curvature.

Let us also remark that, in the context of the heat-kernel techniques,  the Barvinsky--Vilkovisky expansion enables one to obtain the pair creation probability for arbitrary configurations whenever the curvatures are small~\cite{ Akhmedov:2024axn, Boasso:2024ryt}.  Instead, the expressions available in the literature for large curvatures, as far as we know, are considerably more restricted in their range of application to pair creation~\cite{Parker:2009}. The method developed in the following will thus contribute to partially filling this gap, since it is inherently linked to an expansion for strong curvatures and to a nonperturbative effect of pair creation, to which we are going to generically refer as ``Schwinger effect.''

As a last comment before the computations, note that we are going to consider \colm{either Riemannian or} pseudo-Riemannian metrics with a mostly minus signature; we will define the Riemann tensor from the Christoffel symbols $\Gamma^{\mu}{}_{\nu\rho}$ as $R^{\mu}{}_{\nu\alpha\beta}:=\partial_\beta \Gamma^{\mu}{}_{\nu\alpha}+\cdots$, while the Ricci tensor corresponds to the contraction $R_{\mu\nu}:=R^{\rho}{}_{\mu\rho\nu}$.

%%%%%%%%%%%%%%%%%%%%%%%%%%%%%%%%%%%%
%%%%%%%%%%%%%%%%%%%%%%%%%%%%%%%%%%%%
%%%%%%%%%%%%%%%%%%%%%%%%%%%%%%%%%%%%
%%%%%%%%%%%%%%%%%%%%%%%%%%%%%%%%%%%%
%%%%%%%%%%%%%%%%%%%%%%%%%%%%%%%%%%%%

\section{Quantum scalar field in conformally flat metrics}
 \label{sec:scalar_field}
Let us start by considering an arbitrary $\dime$-dimensional Lorentzian spacetime, with a given  metric $g_{\mu\nu}$.
In this universe, define the action for a free quantum, massive, real scalar field $\phi$; it can be written as
\begin{align}
S:= \frac{1}{2}\int \dxd[x][\dime] \sqrt{|g|} \left[ (\nabla\phi)^2 -(m^2+\xi R)\phi^2\right]\, ,
\end{align}
where we have introduced a nonminimal coupling to the Ricci scalar $R$ (with coefficient $\xi$), $m$ is the mass of the field, $\nabla$ is the covariant derivative compatible with the metric and, as customarily, $g$ stands for the determinant of $g_{\mu\nu}$.

In the following we are going to focus on conformally flat spacetimes, whose line elements can be described in terms of a single scalar function $\Omega$.
In conformal time $\tau$, the line element can be written as
\begin{align}
\mathrm{d} s^2=\Omega^2(\tau,{\bf x})\left(\mathrm{d} \tau^2-\mathrm{d} {\bf x}^2\right)\,,
\end{align}
 where we have split the time component, $\tau$, from  the $(\dime-1)$ spatial ones, ${\bf x}$. Importantly, this class of metrics encompasses the FLRW universes, for which $\Omega$ becomes a time-dependent scale factor $a(\tau)$. This can also be particularly useful for $\dime=2$, since in two dimensions every manifold is conformally flat.

Introducing also a Weyl rescaled  field, $\varphi:= \Omega^{(\dime-2)/2}\phi$, the classical action reads
\begin{equation}\label{eq:action_varphi}
    S[\varphi]=\frac{1}{2}\int \!{\rm d}\tau {\rm d}^{(\dime-1)}{\bf x} \left[(\partial\varphi)^2-\Omega^2 \big(m^2 +(\xi-\xi_d)R\big)\varphi^2\right ]  ,
\end{equation}
where $\partial_\mu$ denotes partial derivatives with respect to $\tau$ and $\bf x$, $\xi_\dime:=\frac{(\dime-2)}{4(\dime-1)}$ is the so-called conformal value for the nonminimal coupling  and
\begin{equation}\label{eq:ROmegaInOmega}
R\, \Omega^2 = 2 (\dime-1) \Omega^{-1}\partial^2\Omega
 + (\dime-1)(\dime-4)\Omega^{-2}(\partial\Omega)^2\,.
 \end{equation}

 Observing Eq.~\eqref{eq:action_varphi}, one immediately recognizes that the action corresponds thus to that of a scalar field in Minkowski spacetime with a Yukawa potential
\begin{align}\label{eq:V}
V(\tau,{\bf x}):=m^2 \Omega^2 +\left(\xi-\xi_\dime\right )R\, \Omega^2\, .
\end{align}
To analyze the quantum effects arising in this model, we can, therefore, consider the effective action $\Gamma$ for this field, given by 
\begin{align}\label{eq:EA}
 \Gamma = S + \colm{\Gamma_1, \qquad \Gamma_1= \frac{\mathi}{2} \operatorname{log}\operatorname{Det} \ope\,,}
\end{align}
where we have defined the so-called operator of quantum fluctuations,
\begin{align}\label{eq:def_ope}
 \ope:= \partial^2 +\Omega^2\left[m^2 +(\xi-\xi_\dime)  R\right] \, .
\end{align}

\colm{
For the attentive reader, two technical comments are in order. First, in the corresponding functional integration, it has been shown in Ref.~\cite{Toms:1986sh}  that it is possible  to define a measure in field space that is  invariant under general coordinate transformations using a field variable $\varphi_\omega(x) =|g(x)|^{-\omega/2}\phi(x)$, with an arbitrary weight $\omega$. This is fundamental, for example, to guarantee that the corresponding energy-momentum tensor is conserved.
Alternatively, one can impose a finite renormalization on the energy-momentum tensor, as in Refs.~\cite{Hollands:2004yh, Ferrero:2023unz}.
}

\colm{Second, although the theory defined by Eq.~\eqref{eq:action_varphi} can be quantized using the standard formalism for a scalar field with a spacetime-dependent mass in Minkowski spacetime, if one desires to preserve diffeomorphism invariance, one can devise a renormalization procedure according to one standard prescription of quantum field theory in curved spacetime: in practice, this  amounts to subtracting the ultraviolet divergences associated with the necessary Schwinger--DeWitt coefficients in the heat-kernel expansion~\cite{Birrell:1982} (equivalently, the adiabatic propagator defined on a general curved background).}

\subsection{Resummed heat-kernel approach}\label{sec:hk}
Eq.~\eqref{eq:def_ope} implies that
we can study the effective action of the system in the language of Ref.~\cite{Franchino-Vinas:2023wea}.
We can thus define the heat kernel $K$ associated to the \colm{Euclidean version of } $\ope$ as the solution of the equations
\begin{align}\label{eq:HK_eq}
 [\partial_s - \ope ] K(x,x';s)&=0\,, \quad  K(x,x',0^+)= \delta(x-x')\,,
 \end{align}
where $s$ is an auxiliary parameter which is called the propertime.

 Important to our discussion, Ref.~\cite{Franchino-Vinas:2023wea} has proved that the heat kernel for this type of operator admits a resummed expression from which nonperturbative aspects of pair creation can be studied, with the only assumption that the potential is intense enough; in particular, it could depend on both time and space coordinates.  Defining $\gamma^2_{\alpha\beta}:=2V_{,\alpha\beta}$, the diagonal of the heat kernel takes the form
 \begin{widetext}
\begin{align}\label{eq:HK_diagonal_plb}
K(x,x;s)&=\frac{1}{(4\pi s)^{\dime/2}}\frac{e^{-s V +\partial^{\alpha }V \left[ \gamma^{-3} \left({\gamma s - 2 \tanh\left({\gamma s/2} \right)}\right)\right]_{\alpha \beta } \partial^{\beta }V }}{{\det} ^{1/2}\big((\gamma s)^{-1} \sinh(\gamma s)\big) }  W(x,x;s)\,,
\end{align}
\end{widetext}
where the prefactor  \textit{resums all the invariants} built from $V$, $V_{,\alpha}$ and $V_{,\alpha\beta}$. On the other hand, $W$ contains the information on higher derivatives; in particular, if higher derivatives vanish, $W(x,x;s)=1$. \colm{It is worth emphasizing that substituting the potential from Eq.~\eqref{eq:V} into Eq. \eqref{eq:HK_diagonal_plb} leads to a resummed expression for the heat kernel, which is valid for arbitrary conformally flat metrics, exact for quadratic potentials, and a good approximation whenever derivatives higher than two of the scale factor can be neglected. However, a recast of the resummed expression in terms of geometric invariants is hindered by the fact that general covariance is no longer manifest after choosing conformal coordinates and working with the Weyl-rescaled field.}

\colm{Nevertheless, the geometric interpretation of the expression can be made more transparent as follows. If one considers the resummation of powers of $V$ alone, 
\begin{align}\label{eq:HK_diagonal_potential}
K(x,x;s)&=:\frac{e^{-s V }}{(4\pi s)^{\dime/2}}{}  \tilde W(x,x;s)\,,
\end{align}
which defines the kernel $\tilde W$ on the RHS, one can perform a rescaling in the propertime integral to obtain the following expression for the effective action, 
\begin{align}
\begin{split}
\Gamma_1
=\!-\!\int_0^\infty\!\!\frac{{\rm d}s}{s}
\int \!{\rm d}^\dime x\,  \frac{\sqrt{|g|}e^{-(m^2+(\xi-\xi_\dime)R) s}}{2(4\pi s)^{\dime/2}} \tilde W(x,x;s/\Omega^2)\,,
\end{split}
\end{align}
where we have used $|\Omega|^d=\sqrt{|g|}$. In ($\dime=4$), this form reduces to the Parker--Toms resummation of the heat kernel~\cite{Parker:1984dj}, here adapted to conformally flat spacetimes. ~Eq.~\eqref{eq:HK_diagonal_plb} can then be viewed as a further resummation in this conformal-frame representation, since it also resums derivative terms of $\Omega$ beyond those included in the Parker--Toms result.
}

To further simplify the discussion and the subsequent comparison with Bogoliubov methods, consider a radiation-dominated universe, which corresponds to a metric whose conformal factor is a linear function of conformal time\footnote{In the computation of particle production, an extension of the $\tau>0$  universe to negative conformal times is habitually performed. In our case, we are extending the radiation-dominated universe in a mirror-like way; the interested reader can find in Refs.~\cite{Audretsch:1978qu,Boyle:2018rgh} other alternatives that have been discussed in the literature. \colm{This choice defines an in/out problem on the analytically extended background, directly analogous to the constant-field Schwinger setup.}}
\begin{align}
m^2\Omega^2(\tau,\mathbf x)\rightarrow m^2a^2(\tau) ={b_0^2}{} \tau^2  \,, \quad -\infty < \tau <\infty \,.
\end{align}
For the moment, we will also set $\xi =\xi_\dime$.
From Eq.~\eqref{eq:HK_diagonal_plb}
we thus obtain an \emph{exact} result for the diagonal of the heat kernel, which
in Lorentzian signature\footnote{The heat-kernel expansions obtained in Ref.~\cite{Franchino-Vinas:2023wea} are derived for a Euclidean spacetime. The Lorentzian counterpart is obtained by performing a Wick rotation, which is assumed in the following, keeping a small imaginary part in the propertime $s$.}  reads
\begin{align}\label{eq:HK_diagonal}
K(x,x;s)&=\frac{1}{(4\pi s)^{\dime/2}} \sqrt{\frac{b_0 s}{\cos(b_0 s)\sin(b_0 s)}} e^{-b_0 \tau^2 \tan(b_0 s) } \, .
\end{align}
Importantly, the effective action can be directly computed from a propertime integral of the trace of the heat kernel; in our case, this reduces to computing
\begin{align}
\begin{split}\label{eq:EA_HK}
\colm{ \Gamma_1}&=\colm{\frac{1}{2}}\int \frac{\dd^{\dime}x}{(4\pi)^{\dime/2}}\int_0^{\infty}\frac{\dd s}{s^{1+\frac{\dime}{2}}} \frac{\sqrt{b_0 s}\,e^{-b_0 \tau^2\tan(b_0 s)}}{\sqrt{\cos(b_0 s)\sin(b_0 s)}}\,
\\
&=\colm{\frac{1}{2}} \frac{ \sqrt{\pi} V_0}{(4\pi)^{\dime/2}}\int_0^{\infty}\,\frac{\dd s }{s^{\frac{\dime +1}{2}}} \frac{1}{\sin(b_0 s)}\,,
\end{split}
\end{align}
where $V_0$ has been introduced to denote the spatial volume of the manifold.

Actually, the integral in Eq.~\eqref{eq:EA_HK} is ill-defined, since its integrand has an infinite number of poles, which are placed at $s=\pi n/b_0$, with $n=0,1,\cdots$.
An appropriate way to give meaning to it is to recall that it comes from a Wick rotation, which gives a precise prescription to circumvent the poles and generates a nonvanishing imaginary contribution to the effective action.

Let us discuss the different contributions induced by the poles. \colm{The divergences at $s=0$ are ultraviolet divergences and removed by the usual local small-$s$ subtractions, which renormalize local (covariant) background terms. Instead, the finite poles are not affected by this renormalization process; focusing just on the imaginary part of $\Gamma_1$} and recalling the definition in Eq.~\eqref{eq:inout} for the vacuum persistence probability, it can be shown that
\begin{align}\label{eq:pair_creation_HK}
\frac{P}{2}:=\operatorname{Im} \Gamma &=\frac{V_0 }{\colm{4}(2\pi)^{\dime-1}} b_0^{(\dime-1)/2} \sum_{n=1}^\infty \frac{(-1)^{n+1}}{ n^{(\dime+1)/2}} \,,
\end{align}
which can be written in closed form in terms of a Riemann zeta function $\zeta_{R}(\cdot)$:
\begin{align}
\frac{P}{2}&=-\frac{V_0 }{\colm{4}(2\pi)^{\dime-1}} b_0^{(\dime-1)/2} \left(1-2^{(1-\dime)/2}\right)\zeta_{R}\left(\frac{\dime+1}{2}\right)
\,.
\end{align}
It is important to note that $P$ is the total pair creation probability, i.e. it is already integrated over the entire \colm{spacetime. Importantly, the classical action $S$ is real and local and therefore does not contribute to the imaginary part that determines the vacuum persistence probability.}
 The result in Eq.~\eqref{eq:pair_creation_HK}, evaluated for $\dime=4$, agrees for example with Ref.~\cite{Haouat:2011uy}; the agreement with the Bogoliubov method in an arbitrary number of dimensions will be shown in the following section.

 %%%%%%%%%%%%%%%%%%%%%%%%%%%%%
%%%%%%%%%%%%%%%%%%%%%%%%%%%%%
%%%%%%%%%%%%%%%%%%%%%%%%%%%%%
%%%%%%%%%%%%%%%%%%%%%%%%%%%%%

\subsection{Bogoliubov coefficients}\label{sec:bogoliubov}

In the simple case that we have considered, one can cross check the obtained result by computing the Bogoliubov coefficients linking the in and out vacua. A comprehensive review of this method can be found, for example, in Refs.~\cite{Birrell:1982, Parker:2009}. The first step consists in expanding the quantized field in Fourier modes, which are adapted to the underlying homogeneity of the metric in the spatial coordinates:
\begin{align}
\varphi(\tau, {\bf x})=\int \frac{\mathrm{d}^{\dime-1} {\bf k}}{\sqrt{2(2 \pi)^{\dime-1}}}\left(B_{{\bf k}} e^{\mathi {\bf k} {\bf x}} \varphi_{\bf k}(\tau)+B_{{\bf k}}^{\dagger} e^{-\mathi {\bf k} {\bf x}} \varphi_{\bf k}^*(\tau)\right)\,.
\end{align}
In doing so, we have introduced the creation and annihilation operators, $B_{\bf k}^\dagger$ and $B_{\bf k}$, which satisfy the canonical commutation relations as a consequence of the commutators between $\varphi$ and its conjugate momentum.
 The field equation for the Weyl-rescaled modes can be directly derived from Eq.~\eqref{eq:action_varphi},
\begin{align}\label{eq:EOM_FLRW}
\varphi''_{\bf k}+\omega_{\bf k}^2\varphi_{\bf k}=0\,, \quad \omega_{\bf k}^2:=\mom^2+m^2 a^2+\left(\xi-\xi_\dime\right )Ra^2\,,
\end{align}
where a prime denotes derivatives with respect to the conformal time and
 their corresponding normalization condition is given by
\begin{align}
\varphi_{\bf k} \varphi_{\bf k}^{\prime *}-\varphi_{\bf k}^{\prime} \varphi_{\bf k}^*=2 \mathi\, .
\end{align}

The general solution to Eq.~\eqref{eq:EOM_FLRW}, for $\xi=\xi_\dime$, can be compactly expressed as
\begin{align}\label{eq:solution_modes}
\varphi_{\bf k}(\tau)=C_{{\bf k}, 1} S_{\bf k}(\tau)+C_{{\bf k}, 2} S_{\bf k}^*(-\tau)\,,
\end{align}
where the function $S_{\bf k}$ is essentially a  parabolic cylinder function $D_\nu(z)$,
\begin{align}
S_{\bf k}(\tau):=\left(\frac{2}{ b_0}\right)^{1 / 4} D_{-\frac{1}{2}-2 \mathi \kappa}\left(e^{\mathi \frac{\pi}{4}} \sqrt{2 b_0} \tau\right)\, ,
\end{align}
and we have introduced the rescaled, dimensionless squared momentum $\kappa:=\frac{{\bf k}^2}{4 b_0}$.

In this context, we can naturally define the vacuum states at $\tau \to \pm \infty$. Indeed, for early or late times,  the expansion of the universe slows down and one can naturally define an (infinite order) adiabatic vacuum $\left|0_{\pm}\right\rangle$. First of all, in the late asymptotic region $(\tau \rightarrow +\infty)$ and according to the adiabatic choice,  the preferred (positive-frequency) solution for the late-time modes $\varphi_{\bf k}^{(+)}$ reads~\cite{Boyle:2018rgh,Nadal-Gisbert:2023pum}%\cite{Parker:2009}
\begin{align}
\varphi_{\bf k}^{(+)}(\tau \rightarrow +\infty) \overset{}{ \sim} \frac{e^{-\mathi \int_\tau \omega_{\bf k}(u) {\rm d} u}}{\sqrt{\omega_{\bf k}(\tau)}} \sim \frac{e^{-\mathi\frac{b_0}{2} \tau^2-\mathi\kappa \ln \left(2 b_0 \tau^2\right)}}{\sqrt{b_0 \tau}}\, ,
\end{align}
where $\omega_{\bf k}(\tau)=\sqrt{{\bf k}^2+m^2 a^2(\tau)}$.  If we impose this late-time behavior in the general solution~\eqref{eq:solution_modes}, the coefficients are completely determined to be
\begin{align}
C^+_{{\bf k}, 1}= e^{-\frac{\pi \kappa}{2}+\mathi \frac{\pi}{8}}, \quad C^+_{{\bf k}, 2}=0\, .
\end{align}
Afterwards, expanding the field in terms of $\varphi_{\bf k}^{(+)}$ and its complex conjugate, we can define the Fock space corresponding to the accompanying annihilation and creation operators, respectively $B^{(+)}_{{\bf k}}$ and $B^{(+),\dagger}_{{\bf k}}$, with the vacuum $\vert 0_+\rangle$ defined as the state containing no $B^{(+)}_{\bf k}$ particles.

 Analogously, the early-time adiabatic vacuum $\left|0_{-}\right\rangle$ is determined by the adiabatic early-time modes $\varphi_{\bf k}^{(-)}$, which in turn satisfy the asymptotic condition
\begin{align}\label{eq:phi_minus_asym}
\varphi_{\bf k}^{(-)}(\tau \rightarrow-\infty) \sim \frac{e^{-\mathi \int_\tau \omega_{\bf k}(u) d u}}{\sqrt{\omega_{\bf k}(\tau)}} \sim \frac{e^{\mathi\frac{b_0}{2} \tau^2+\mathi\kappa \log \left(2 b_0 \tau^2\right)}}{\sqrt{-b_0 \tau}}\, ;
\end{align}
a solution satisfying such constraint is obtained from Eq.~\eqref{eq:solution_modes} by choosing the following coefficients:
\begin{align}
 C^-_{{\bf k}, 1}=0\,, \quad C^-_{{\bf k}, 2}=e^{-\frac{\pi \kappa}{2}-\mathi \frac{\pi}{8}}\, .
\end{align}

At late times, we can expand the modes $\varphi_{\bf k}^{(-)}$ in terms of $\varphi_{\bf k}^{(+)}$ and its conjugate, since they form a basis:
\begin{align}
\varphi_{\bf k}^{(-)}(\tau \to \infty)&=\alpha_{\bf k}\varphi_{\bf k}^{(+)}(\tau)+\beta_{\bf k} \left(\varphi_{\bf k}^{(+)}\right)^*(\tau)\, .
\end{align}
This relation tightly links both types of modes, being $\alpha_{\bf k}$ and $\beta_{\bf k}$ the so-called  Bogoliubov coefficients, which explicitly  depend on the momenta of the modes involved. For the present spacetime, we find
\begin{align}
\alpha_{\bf k}= \frac{e^{-3\pi \kappa}}{\sqrt{2 \pi}} \Gamma\left(\tfrac{1}{2}+2\mathi\kappa\right) \left(1+e^{4\pi \kappa}\right)\, , \;\; \beta_{\bf k}=-\mathi e^{-2\pi \kappa} \, ,
\end{align}
which  satisfy the Bogoliubov consistency condition $\vert \alpha_{\bf k}\vert^2-\vert \beta_{\bf k}\vert^2=1$.

Now let us prepare our system such that at early times $\vert\text{in}\rangle=\vert 0_-\rangle$. At late times, an observer will naturally define particles through the late-time Fock space, i.e. $\vert\text{out}\rangle=\vert 0_+\rangle$, so the vacuum persistence probability corresponds to\footnote{\colm{Note that, because of the symmetry in the pair of created particles for a real scalar field, the integral in momentum space should be done just over half of the space~\cite{Parker:2009}}.}
\begin{align}
\big\vert \langle 0_- \vert 0_+\rangle \big\vert^2=\exp \left[-{\frac{V_0}{\colm{2}}}\int \frac{{\rm d}^{\dime-1} {\bf k}}{(2\pi)^{\dime-1}} \log \left|\alpha_{\bf k}\right|^2\right] \, ,
\end{align}
where it should be recalled that $V_0$ is the spatial volume. One can readily compute $\vert\alpha_{\bf k}\vert^2=1+e^{-4 \pi \kappa}$, either from its definition or from the Bogoliubov consistency  relation; matching to Eq.~\eqref{eq:inout}, the probability of pair creation can be seen to be
in agreement with Eq.~\eqref{eq:pair_creation_HK}.

At this point we can make an  interesting remark related to  the pair production rate in Eq.~\eqref{eq:pair_creation_HK}. This formula is similar but not equal to Schwinger's result for a constant and homogeneous electric background $E$ in SQED; indeed, for massive fields of mass $m_{\rm SQED}$ and charge $e$, it establishes that the pair creation probability integrated during a period of time $T_0$ is
\begin{equation}
    \frac{P_{\rm SQED}}{V_0}=2\pi T_0\left(\frac{eE}{4\pi^2}\right)^{\dime/2} \sum_{n=1}^{\infty} \frac{(-1)^{n+1}}{ n^{\dime/2}} e^{-n\pi m_{\rm SQED}^2/(eE)}\, .
\end{equation}

 The reason for such a difference can be traced back, for example, to the equation for the modes in the Bogoliubov approach.
In  SQED the corresponding frequencies are
$\omega_{\vec k,{\rm SQED}}^2= m_{\rm SQED}^2 + k_\perp ^2+ (k_\parallel+eE_0 t)^2$, where we have split the momentum components according to whether they are parallel ($k_\parallel$) or  perpendicular $(k_\perp)$ to the polarization of the electric field.
On the other hand, for a radiation-dominated universe the frequencies are
$\omega^2_{\vec k,{\rm rad}}=k_\perp^2+k_\parallel^2+b_0^2\tau^2$.
This implies that the corresponding Bogoliubov coefficients are the same only when $k _\parallel=0$ and $m_{\rm SQED}=0$, as long as we identify $eE\equiv b_0$;
additionally, in SQED, the integration over $k_{\parallel}$ gives the time interval and not a further power of $n$.

It is worth noticing that, contrary to the situation in the Schwinger effect, the pair creation probability in a radiation-dominated universe does not display an exponential suppression with the mass, even if the field is massive.
%%%%%%%%%%%%%%%%%%%%%%%%%%%%%%%%%%%%%%
%%%%%%%%%%%%%%%%%%%%%%%%%%%%%%%%%%%%%%
%%%%%%%%%%%%%%%%%%%%%%%%%%%%%%%%%%%%%%
%%%%%%%%%%%%%%%%%%%%%%%%%%%%%%%%%%%%%%

\section{Other Gravitational analogues}\label{sec:analogues}

Our heat kernel master formula~\eqref{eq:HK_diagonal_plb} is rather versatile, since it is only linked to the operator $\ope$ in Eq.~\eqref{eq:def_ope}. Indeed, we have already shown that results in a gravitational setup can be studied by analyzing an analogous problem with a spacetime-dependent mass.

There is a further type of analogy that is immediate but, somehow, has not been pursued in the literature before: the case in which, for a massless field, the nonminimal coupling to the curvature is responsible for pair creation. We will refer to pair creation in this scenario as curvature-induced; its relevance has been previously considered, for instance, in the context of generation of dark matter during reheating, when $Ra^2$ oscillates~\cite{Cembranos:2019qlm, Fairbairn:2018bsw,Markkanen:2015xuw}.

The condition to find a curvature analogue of the Schwinger effect  is 
\begin{align} \label{eq:Rcondition}
R a^2\propto \tau^2+ c\,,
\end{align}
with $c$ a real constant.  \colm{Note that, for \(c=0\), the term $Ra^2$ gives a contribution which is equivalent to the mass term $m^2a^2$ in the radiation-dominated case discussed in Sec.~\ref{sec:hk}. Therefore, the model leads to the same results in terms of modes and pair creation probability}. On the other hand, as we are going to see in Case II below, a non-vanishing $c$ provides a \colm{further  analogy} in which $c$ plays the role of a mass term in the equation for the modes $\varphi_{\bf k}$. Let us analyze these alternatives in detail.

\textbf{Case I.} \colm{In $\dime=4$, Eq.~\eqref{eq:ROmegaInOmega} gets simplified, since the last factor trivially vanishes; if on top of that $c=0$, we can determine the analogous scale} factor by solving the differential equation
\begin{equation}
(\xi-\xi_4) R a^2 =6(\xi-\xi_4)   \frac{a''}{a} =: 6 (\xi-\xi_4)b_0^2 \tau^2 \, ,
\end{equation}
where we inserted a factor 
 $6(\xi-\xi_4)$ into the last equality for convenience. The general solution to this differential equation can be obtained as a special case of Eq.~\eqref{eq:solution_modes}:
\begin{equation}
a(\tau)=c_-  D_{-1/2}(\sqrt{2b_0} \tau)+c_+ D_{-1/2}(-\sqrt{2b_0} \tau)\, .
\end{equation}
The function $D_{-1/2}(-\sqrt{2b_0}\tau)$
is a positive and monotonically increasing function, whose asymptotic expansion for large $\tau>0$ can be read from Eq.~\eqref{eq:phi_minus_asym}.
In the general case where both coefficients $c_{\pm}$ are \colm{positive}, the scale factor describes a bouncing universe. In cosmological time, defined as $t:=\int_0^{\tau} a(\tau_1) {\rm d}\tau_1$, one can obtain the large-time behaviour of the  scale factor, $a(t)\approx \sqrt{2 b_0} t \sqrt{\log (\sqrt{b_0}t)}$. If instead one of the coefficients $c_{\pm}$ vanishes, the universe either collapses or shows a big bang, both at a finite cosmological time.
\colm{On the other hand, if the coefficients $c_\pm$ have opposite signs, the scale factor vanishes at a finite conformal time: near this time, the universe behaves as  a radiation-dominated one.}

As already said, in this case the pair creation probability can be obtained from Eq.~\eqref{eq:pair_creation_HK} by a simple rescaling of $b_0$. Curiously, the parabolic cylinder functions play a dual role here: on the one hand, they determine the function $a(\tau)$, while, on the other hand, they are crucial to compute the modes of the field.

\textbf{Case II.} \colm{In an arbitrary number of dimensions other than four, the equation determining the analogous scale factor becomes nonlinear due to the last term in Eq.~\eqref{eq:ROmegaInOmega}, implying a higher level of intricacy in obtaining its solutions.  Still, we will be able to find a simple model whose vacuum persistence probability is  no longer described by expression~\eqref{eq:pair_creation_HK}. Consider} a situation in which
$Ra^2$ is a quadratic polynomial  in $\tau$. A simple choice in an arbitrary number of dimensions $\dime>2$ is a Gaussian scale factor, $a(\tau)=a_0\exp{-\alpha\tau^2/2}$, where $a_0$ and $\alpha$ are real parameters that respectively govern its intensity and its time dependence. For this simple geometry, the Ricci scalar is given by
\begin{align}\label{curv}
    R&= \frac{2(\dime-1) \alpha}{a_0^2}  \left(-1+\frac{\dime-2}{2} \alpha  \tau^2\right) e^{\alpha \tau^2}\,,
\end{align}
so that $Ra^2(\tau)={2 (\dime-1) \alpha}  \left(\frac{\dime-2}{2}\alpha  \tau^2-1\right)$.

Note that, in order to apply our methods, the coupling to the curvature should be such that $\xi-\xi_\dime>0$, so that the effective potential is confining (and not unstable) for large values of  $\tau$; we shall also restrict $\alpha\leq0$, so that the quantum field is not tachyonic.
In this universe, where the cosmological time can be written in terms of the imaginary error function $\operatorname{Erfi}$,
\begin{align}
  \frac{t}{a_0}=\left( \frac{\pi}{2|\alpha|}\right)^{1/2} \operatorname{Erfi}\left(\sqrt{\frac{|\alpha|}{2}}\tau\right),\quad t\in\mathbb{R}\,,
\end{align}
the scale factor describes once again a bouncing universe: it reaches a minimum $a(\tau=0)=a_0$ and expands slower than an exponential.
Indeed, using the asymptotic expansion for the imaginary error function, one can show that, in cosmological time, 
 $a(t)\simeq \sqrt{2\vert\alpha\vert} t
\sqrt{ \log(\sqrt{\vert\alpha\vert}t)} $ for large values of $t$, which is essentially the same behavior found in the previous Case I. 
For the sake of completeness, since, to the best of our knowledge, this universe has not been widely studied in the literature, \colm{let us give an effective perfect-fluid interpretation. In four dimensions, and assuming a perfect fluid source, Einstein's equations give $R a^2 =\kappa a^2 (\rho-3p)$, where $\kappa$ is Einstein's gravitational constant. This expression implies  that the condition in Eq.~\eqref{eq:Rcondition} fixes the trace of the stress-energy tensor, up to a factor of $a^2$. Equivalently, we can introduce an effective equation of state parameter 
\begin{equation}
w_{\text{eff}}:=\frac{p}{\rho}=\frac{1}{3}-\frac{Ra^2}{9\mathcal{H}^2}\, ,\qquad \mathcal{H}:=\frac{a'}{a}\, .
\end{equation} 
From the previous definition, one can see that determining the energy conditions requires not only the constraint in Eq.~\eqref{eq:Rcondition}, but also a  solution for $a(\tau)$. Moreover, it is well known that the strong energy condition is violated if $w_{\text{eff}}<-1/3$. Taking all this into account, for Case II one finds that the strong energy condition is violated  for all $\tau$, while the null, weak, and dominant energy conditions are violated in the region $\tau^2<1/|\alpha|$.
}

The diagonal of the heat kernel and the effective action corresponding to this universe can be readily computed using Eq.~\eqref{eq:HK_diagonal_plb}. The result is
\begin{align}
\begin{split}\label{eq:EA_HK_curvature}
\colm{\Gamma_1}&= \frac{ \sqrt{\pi} V_0}{\colm{2}(4\pi)^{\dime/2}}\int_0^{\infty}\,\frac{\dd s }{s^{\frac{\dime +1}{2}}} \frac{e^{-\tilde{m}^2 s}}{\sin(\tilde{a} s)}\,,
\end{split}
\end{align}
where the effective mass and frequency are respectively given by
\begin{align}
\tilde{m}^2:&=2 (\dime-1)(\xi-\xi_\dime)|\alpha|\,,\\
\tilde{a}^2:&=(\dime-1)(\dime-2)(\xi-\xi_d)\alpha^2 \,.
\end{align}
In this case, if $\xi>\xi_\dime$, the curvature itself entails an expected exponential suppression in the creation of pairs,
\begin{align}\label{eq:pair_creation_curvature}
\frac{P}{2}=\frac{V_0 }{\colm{4}(2\pi)^{\dime-1}} \tilde{a}^{(\dime-1)/2} \sum_{n=1}^\infty \frac{(-1)^{n+1}}{ n^{(\dime+1)/2}} e^{-\tilde{m}^2 n\pi/\tilde{a}}\,.
\end{align}
From this formula one can readily see that, in the large-$\dime$ limit, the pair creation could be greatly enhanced by the prefactor $\tilde{a}^{(d-1)/2}\sim (d-1)^{(d-1)/2}$, since the quotient $\tilde{m}^2/\tilde{a}\sim d^0$ and thus the exponential counting the number of created pairs does not counteract the effect.

\section{Discussion}\label{sec:discussion}

In this letter we have built on the results of Ref.~\cite{Franchino-Vinas:2023wea}, where resummed effective actions for Yukawa potentials have been obtained, with a two-fold aspiration. 

First, we have examined the applicability of our heat-kernel techniques in a situation other than the well-known homogeneous electric field. As a test-bed, we have analyzed the pair creation probability in a radiation-dominated universe with an arbitrary number of dimensions, which was shown to agree with the computation in terms of the corresponding Bogoliubov coefficients.
This agreement is non-trivial, inasmuch as the heat-kernel derivations are computed in Riemannian signature and require a subsequent Wick rotation; in fact, in spite of recent advances~\cite{Kontsevich:2021dmb}, a complete proof of the mathematical validity of Wick rotations is still lacking. 

Second, we have pointed out the generality of our results. In effect, they are not limited to Yukawa setups: they include electromagnetic and gravitational backgrounds, among them the familiar FLRW metrics. In cases where an intensity scale is dominant, our method can be directly used, without the need to solve an equation for the modes, as would be the case in the Bogoliubov approach.

Note that the range of applicability of our method includes scenarios with arbitrary conformally flat spacetime backgrounds and, thus, it could be used to analyze pair creation even in static universes.  According to our view, however, the only way for such effect to be present, is through an instability at the classical level. This can be intuited, for instance, by analyzing the effective potential arising  for the Schwinger effect in the space-dependent gauge \cite{Padmanabhan:1991uk} or the tachyonic instabilities discussed in Ref.~\cite{Landulfo:2012nz} (see also~\cite{Schroer:1970ddg}).  Extensions to conformal factors that depend on both space and time coordinates, which easily become intractable under the Bogoliubov approach, seem feasible with the aid of Eq.~\eqref{eq:HK_diagonal_plb}; these ideas and possible applications to inflation scenarios~\cite{Domcke:2019qmm, vonEckardstein:2024tix, VicenteGarcia-Consuegra:2025lkh} are being explored.

 Beyond the use of heat kernel techniques, we have highlighted the existence of gravitational analogues of the Schwinger effect for massless,  nonconformally coupled scalar fields, for cosmological evolutions in which $Ra^2$ is a quadratic function of conformal time.  Further examples could be obtained by choosing the scale factor to satisfy the differential equation
\begin{equation}\label{eq:general_tanh}
R a^2 = A + B \tanh(\rho\tau) + C \tanh^2(\rho\tau)\,,
\end{equation}
whose solutions for $a$ can be written in terms of hypergeometric functions~\cite{Birrell:1982, Haouat:2011uy}.

 Coming back to our results for the pair-creation probability, it is curious to observe that the radiation-dominated universe involves an effective horizon, akin to the one discussed in Ref.~\cite{Ilderton:2025umd}; \colm{in our case, it coincides with the singular surface at $\tau=0$. More in detail, a massive particle following a geodesic with a nontrivial momentum would reach the speed of light 
 when extrapolated to $\tau=0$, where $a$ vanishes. This fact can be heuristically understood as an effective separation of regions and may 
 suggest a connection with pair creation. However, backgrounds with such an effective horizon 
 constitute only a subclass of those admitting pair creation;  indeed, it is not hard to see that, for some of the scale factors obtained from Eq.~\eqref{eq:general_tanh}, this does not occur and, nevertheless, multiparticle states are excited. Still, a classification of spacetimes based on this criterion might prove useful and, in our opinion, deserves further inspection.}

In any case, obtaining resummed expressions in a broader class of scenarios is still required. For example, Ref.~\cite{Parker:2009} shows that the Euler--Heisenberg-like heat kernel for massless spinors in gravitational backgrounds,
\be
K_{\rm F}(x, x;\tau)=\mathi\left\{\operatorname{det}\left[\frac{ R_{\mu \nu a b} \Sigma^{a b}}{4 \pi \mathi\sinh \left(\tau R_{\mu \nu a b} \Sigma^{a b}\right)}\right]\right\}^{1 / 2}\, ,
\ee
where $\Sigma_{\alpha \beta} :=\frac{1}{4}\left(\gamma_\alpha \gamma_\beta-\gamma_\beta \gamma_\alpha\right)$ and $\gamma_\mu$ are the Dirac matrices, correctly reproduces the (gravitational) axial anomaly. This suggests a potential resummation candidate for  the contributions arising from the gamma matrices.
Condensed-matter analogues may provide further hints for identifying similar resummation structures.

General expansions  of effective actions  for large curvatures in arbitrary spacetimes are also missing and might shed light on the connection between perturbative approaches and Hawking radiation; work along these lines is currently being pursued.

%%%%%%%%%%%%%%%%%%%%%%%%
%%%%%%%%%%%%%%%%%%%%%%%%
%%%%%%%%%%%%%%%%%%%%%%%%
%%%%%%%%%%%%%%%%%%%%%%%%

\section*{Acknowledgments}
The authors acknowledge useful discussions with C.~García-Pérez and V.~Vitagliano. SAF thanks the members of the Institut Denis Poisson, especially M. Chernodub, for their warm hospitality.  The research activities of SAF and FDM have been carried out in the framework of Project PIP 11220200101426CO, CONICET. SAF acknowledges support from the INFN Research Project QGSKY and Project 11/X748 of UNLP. The authors would like to acknowledge the contribution of the COST Action CA23130.
The authors also extend their appreciation to the Italian National Group of Mathematical Physics (GNFM, INdAM) for its support.  The authors lastly acknowledge fruitful discussions and funding from the workshops ``New Trends in First Quantisation: Field Theory, Gravity and Quantum Computing'' (Heraeus Stiftung).

%%%%%%%%%%%%%%%%%%%%%%%%%%%%%
%%%%%%%%%%%%%%%%%%%%%%%%%%%%%
%%%%%%%%%%%%%%%%%%%%%%%%%%%%%
%%%%%%%%%%%%%%%%%%%%%%%%%%%%%

\bibliography{biblio.bib}
\end{document}